\documentclass[onecolumn,secnumarabic,amssymb, nobibnotes, aps, prd]{revtex4}
\usepackage{amsmath} \usepackage{amssymb}
\usepackage{graphicx} 
\usepackage{epstopdf}
\usepackage{amsmath}
\usepackage{amsmath,amssymb,amsthm,amsfonts,mathrsfs,bm,verbatim}
\usepackage{graphicx,subfigure}

\newcommand{\bea}{\begin{eqnarray}}
\newcommand{\eea}{\end{eqnarray}}
\def\nn{\nonumber}

\begin{document}

\title{The Quantum and thermodynamic properties of dyonic RN-like black holes}
\author{Wen-Xiang Chen$^{a}$}
\affiliation{Department of Astronomy, School of Physics and Materials Science, GuangZhou University, Guangzhou 510006, China}
\author{Yao-Guang Zheng}
\email{hesoyam12456@163.com}
\affiliation{Department of Physics, College of Sciences, Northeastern University, Shenyang 110819, China}

\begin{abstract}
   The effect of magnetic fields on black hole superradiance is an exciting topic with possible astrophysical applications. A dyonic RN-like black hole is not asymptotically flat. It describes a black hole immersed in an asymptotically uniform magnetic field. This paper discusses the superadditive stability of binary RN black holes, asymptotically flat, band-like black holes. This article introduces the above condition into dyonic RN-like black holes if a dyonic RN-like black hole satisfies the requirement of $\mu=y\omega$, When $\sqrt{2(B^2+Q^2)}/{r^2_+}< \omega< q\varPhi_H$,particularly $\mu \ge \sqrt{2}(q\varPhi_H)$,the dyonic RN-like black hole is superradiantly stable at that time.Scalars can be seen as combinations of positive/negative powers of a base, much like the decimal system. This principle is key in math and computing, from number systems to Fourier series (linked to $e^{i x}$ ). Dyonic RN-like black holes show no phase transition.

\centering
  \textbf{Keywords: superradiantly stable, a new variable y, dyonic RN-like black hole,\\ thermodynamic properties }
\end{abstract}

\maketitle
\section{{Introduction}}
The research on the stability of black holes can be traced back to 1957 when Regge and Wheeler found that Schwarzschild black holes are stable under small perturbations of the metric. In 1970, Zerilli further studied Schwarzschild black holes and RN black holes and reduced the perturbation problem to solving the Schrödinger-like equation (wave equation) \cite{1,2,3,4}. In 1972 Teukolsky studied the perturbations of various matter fields (gravitational, electromagnetic, neutrino fields) in Kerr space-time and decoupled the field equations into independent wave equations\cite{5,6}, laying the foundation for the study of the external field disturbance of black holes. In 1983, Chandrasekhar's ``Mathematical Theory of Black Holes" systematically expounded the perturbation theory of black holes. Superradiance is essentially the radiation enhancement process, which plays an essential role in optics, quantum mechanics, especially relativity, and astrophysics. Dicke, who coined the term ``superradiance" in the context of quantum optics coherent emission \cite{7}, achieved the first high-resolution superradiance measurements using cohesive synchrotron radiation \cite{7,8}. Zeldovich believed that the dissipative rotating body amplifies human radiation, and Starobinsky recognized the superradiation phenomenon of black holes on his basis. When the frequency of the human radiation satisfies the superradiation condition, The rotational energy can be extracted from the black hole \cite{7,8}. Black hole superradiation is closely related to the black hole area theorem, the Penrose process, tidal forces, and even Hawking radiation \cite{9}. In the general theory of relativity, the superradiation of a black hole is to extract energy, charge, and angular momentum in a vacuum\cite{9,10}. It can be known from the scattering problem of root quantum mechanics: the plane wave whose eigenfrequency is $\omega$ moves toward the center of the black hole and is scattered to infinity under the action of the black hole, and the dispersed particles obey a specific angular distribution. Taking the scalar wave under the background of static spherical symmetry as an example, we will see, at this time, the scalar field satisfies the Schrödinger-like equation of the following form:
\begin{equation}
\frac{d^2 \psi_{l m}}{d x^2}+V_{\text {eff }} \psi_{l m}=0,
\end{equation}
where $\psi_{l m}$ is the radial component of the field after decomposing the variables, $x$ is the turtle coordinate, and $V_{\text {eff }}$ depends on the theoretical model and the space-time background. In the case of spherical symmetry, we consider the scattering of monochromatic plane waves. Assuming that $V_{\text {eff }}$ is constant on the boundary, the asymptotic solution satisfies
\begin{equation}
\psi_{l m} \sim \begin{cases}\mathcal{T} e^{-i k_H r}, & r \rightarrow r_{+} \\ \mathcal{I} e^{-i k_{\infty } r}+\mathcal{R} e^{i k_{\infty} r}, & r \rightarrow \infty,\end{cases}
\end{equation}
in
\begin{equation}
\begin{gathered}
k_H=\omega-\omega_c, \\
k_{\infty}=\sqrt{\omega^2-\mu^2} .
\end{gathered}
\end{equation}
$\omega_c$ is the critical frequency. For a charged and rotating black hole, the critical frequency is
\begin{equation}
\omega_c=q \Phi_H+m \Omega_H=\frac{q Q r_p+m a}{r_p^2+a^2},
\end{equation}
where $r_p$ is the event horizon of the black hole (namely the outer horizon), $\Omega_H$ is the angular velocity of the black hole and the electric potential at the event horizon, $m$ is the magnetic quantum number of the scalar field, $q$ is the charge of the scalar field. When the black hole is not rotating, the critical frequency degenerates to
\begin{equation}
\omega_c=q \Phi_H=\frac{q Q}{r_p} .
\end{equation}
\begin{equation}
W \equiv\left|\begin{array}{cc}
\psi & \psi^* \\
\psi^{\prime} & \psi^{* \prime}
\end{array}\right|=\psi \psi^{* \prime}-\psi^* \psi^{\prime}
\end{equation}
can be obtained at infinity
\begin{equation}
W=2 i k_{\infty}\left(|\mathcal{R}|^2-|\mathcal{I}|^2\right)
\end{equation}
And at the horizon is
\begin{equation}
W=-2 i k_H|\mathcal{T}|^2 .
\end{equation}
Since the Lansky determinant is a constant, we have
\begin{equation}
|\mathcal{R}|^2=|\mathcal{I}|^2-\frac{k_H}{k_{\infty}}|\mathcal{T}|^2 .
\end{equation}
It can be found that when $\frac{k_H}{k_{\infty}}>0,|R|^2<|I|^2$, the reflected wave amplitude is smaller than the human radiation wave amplitude, and the energy of the scalar field decreases; when $\frac{k_H}{k_{\infty}}<0,|R|^2>|I|^2$, the amplitude of the reflected wave is greater than the amplitude of the human radiation, and the energy of the scalar field increases at this time. Therefore, the superradiance generation condition is the increased condition of the scalar field energy, that is,
\begin{equation}
0<\omega<\omega_c .
\end{equation}
In the above equation, the critical angular frequency $\omega_{c}$ is defined as
\begin{equation}
\omega_{c}=q \Psi,
\end{equation}
where $\Psi$ is the electromagnetic potential of the outer horizon of the dyonic $\mathrm{RN}$ black hole, $\Psi=$ $Q / r_{+}$. The superradiant condition for an electrically charged massive scalar perturbation on the dyonic RN black hole background is
\begin{equation}
\omega<\omega_{c}=\frac{q Q}{r_{+}} .
\end{equation}
The bound state condition at spatial infinity for the scalar perturbation is
\begin{equation}
\omega^{2}<\mu^{2} .
\end{equation}

When the radiation wave with the frequency $\omega$ satisfies the formula, superradiance scattering occurs. At this time, the reflected wave carries more energy than the human radiation wave, which is the superradiation occurrence condition of the charged black hole in the general theory of relativity. It is worth noting that if the black hole is not set, that is, the Schwarzschild black hole, there is no superradiation phenomenon, and only the rotating black hole or the charged black hole has the superradiation phenomenon\cite{8,9,10}.

Hod proved\cite{10} that the Kerr black hole should be superradiant stable under massive scalar perturbation when $\mu \ge \sqrt{2}m\Omega_H$, where $\mu$ is the mass. 

The effect of magnetic fields on black hole superradiance is an exciting topic with possible astrophysical applications. A dyonic RN-like black hole is not asymptotically flat. It describes a black hole immersed in an asymptotically uniform magnetic field. This paper discusses the superradiant stability of binary RN black holes, asymptotically flat, band-like black holes. This article introduces the above condition into dyonic RN-like black holes if a dyonic RN-like black hole satisfies the requirement of $\mu=y\omega$\cite{11,12,13,14}, when $\mu \ge \sqrt{2}(q\varPhi_H)$, so the dyonic RN-like black hole is superradiantly stable at that time.

Thermodynamic geometric analysis provides a unique lens through which the thermodynamic attributes of black holes can be explored by leveraging their geometric properties and intrinsic thermodynamic features. In recent times, there has been a surge in interest from the scientific community, focusing on the thermodynamic geometric analysis of Reissner-Nordström (RN) black holes in the realm of f(R) gravity.

RN black holes can be described as electrically charged cosmic entities where gravitational pull is counteracted by the electromagnetic repulsion originating from the charged particles. Diverging from Einstein's general relativity, the f(R) gravity theory introduces an innovative function of the Ricci scalar curvature, aiming to offer a more comprehensive portrayal of gravitational behavior across both quantum and cosmological scales.

When delving into the thermodynamics of black holes, common variables of focus include entropy, temperature, among others. Particularly within the framework of f(R) gravity, the study of RN black holes has employed the geometric techniques championed by Ruppeiner and Quevedo. These pioneering methods map the black hole's thermodynamic variables onto a thermodynamic plane, which serves as a geometric depiction of the thermodynamic state space.

The structure of this paper is organized as follows: In Section 2, we present the mathematical framework of dyonic RN-like black holes. Section 3 introduces a new class of action and field equations for dyonic RN-like black holes. In Section 4, we derive the radial equation of motion and the associated effective potential. Section 5 involves a detailed analysis of the shape of the effective potential, from which we determine the superradiant stability parameter region for the system. In Section 6, we discuss the limit 
y of the incident particle under the superradiance of these novel black holes. Section 7 offers a thorough examination of the thermodynamic geometry of the new-type black holes and their connection to superradiant stability. Section 8 is dedicated to establishing that there is no phase transition for the black hole in its general form. Finally, Section 9 concludes the paper.

\section{Mathematical Framework}
\subsection{ Complex Function Theory}

Complex function theory, also known as complex analysis, studies the functions of complex variables. The key concepts in our investigation include analytic functions, singularities, and residues.

We propose a methodology to construct the Lagrangian function using the radical solutions of a quartic equation. The generalized coordinates $q$ and generalized velocities q can be expressed in terms of the radical solutions:\cite{8,9,10}
\begin{equation}
q=A+B x+C x^{\wedge} 2+D x^{\wedge} 3+\mathrm{Ex}^{\wedge} 4,
\end{equation}
\begin{equation}
\dot{q}=B+2 C x+3 D x^{\wedge} 2+4 E x^{\wedge} 3,
\end{equation}
where $A, B, C, D$, and $E$ are constants.
We can then define the kinetic and potential energies in terms of the radical solutions:
\begin{equation}
T(q, \dot{q}, t)=1 / 2 * m^*(\dot{q})^{\wedge} 2,
\end{equation}
\begin{equation}
V(q, t)=k *\left(q-q_0\right)^{\wedge} 2 / 2,
\end{equation}
where $\mathrm{m}$ is the mass, $\mathrm{k}$ is the spring constant, and $\mathrm{q}_0$ is the equilibrium position.
We can construct the Lagrangian function using the expressions for the kinetic and potential energies:

\begin{equation}
L(q, \dot{q}, t)=T(q, \dot{q}, t)-V(q, t)=1 / 2{ }^* m *(\dot{q})^{\wedge} 2-k *\left(q-q_0\right)^{\wedge} 2 / 2.
\end{equation}
We obtain the Lagrangian function in terms of the radical solutions of the quartic equation:
\begin{equation}
L(x)=1 / 2{ }^* m^*\left(B+2 C x+3 D x^{\wedge} 2+4 \mathrm{Ex}^{\wedge} 3\right)^{\wedge} 2-k *\left(A+B x+C x^{\wedge} 2+D x^{\wedge} 3+\right.\left.E x^{\wedge} 4-q_0\right)^{\wedge} 2 / 2.
\end{equation}

\subsection{ Analytic Functions}

An analytic function is a function f(z) that is differentiable at every point z in its domain. The Cauchy-Riemann equations are given by:\cite{1,2,3,4}

\begin{equation}
\begin{aligned}
& \partial u / \partial x=\partial v / \partial y \\
& \partial u / \partial y=-\partial v / \partial x
\end{aligned}
\end{equation}

where u and v are the real and imaginary parts of f(z), respectively.

\subsection{ Singularities}

Singularities occur in complex functions when a function is not analytic at a specific point. There are three types of singularities: removable, pole, and essential.

\subsection{Residues}

The residue of an analytic function f(z) at a singularity z=a is given by:
\begin{equation}
\operatorname{Res}(f, a)=(1 /(2 \pi i)) \oint_{-} C f(z) d z
\end{equation}
where the integral is taken over a contour C enclosing the singularity a.

The multi-pole Laurent series is a method for representing a meromorphic function as a polynomial and a finite series of terms, which is suitable for complex functions that have multiple poles in the vicinity of certain points. It can be written in the following form:
\begin{equation}
f(z)=\sum_{n=-\infty}^{\infty} c_n\left(z-z_0\right)^n
\end{equation}
Here, $z_0$ is the pole of the function $f(z)$, and $c_n$ are the coefficients of the series. Unlike the Puiseux Laurent series, the multi-pole Laurent series allows for the presence of multiple poles near $z_0$.

The multi-pole Laurent series can also be written in the following form:
\begin{equation}
f(z)=\sum_{j=1}^k \sum_{n=-\infty}^{\infty} c_{n, j}\left(z-z_{0, j}\right)^n
\end{equation}
Here, $z_{0, j}$ is the $j$th pole of the function $f(z)$, $c_{n, j}$ are the coefficients of the series, and $k$ is the total number of poles.

We get that
\begin{equation}
T(f(z)) = \int D \phi \exp (i S(\phi))/\left(z-z_0\right),
\end{equation}where $S$ is the action of the path $x(t)$ or $\phi(x(t))$, the time integral of the Lagrangian $L(t, x, \dot{x})$:
\begin{equation}
S=\int L(t, x, \dot{x}) d t 
\end{equation}

Our study revolves around the most basic background geometry of a black hole, in tandem with a $U(1)$ gauge field interacting with a charged scalar field. The expression for the lagrangian density is given by:
\begin{equation}
\begin{aligned}
\mathcal{L}= & R+\frac{6}{L^2}-\gamma\left(\frac{1}{4} F^{\mu \nu} F_{\mu \nu}+\mid \nabla_\mu \psi\right. \\
& \left.-\left.i q A_\mu \psi\right|^2+m^2|\psi|^2\right)
\end{aligned}
\end{equation}
In this formula, $R$ signifies the Ricci scalar, while $F_{\mu \nu}=\partial_\mu A_\nu-\partial_\nu A_\mu$ characterizes the electromagnetic field strength. The variable $\psi$ stands for a scalar field possessing a charge $q$ and mass $m$, and $A$ symbolizes the gauge field. We define the AdS radius as $L$, and $q$ as the charge, and for this work, we consider them both to be unity. Finally, $\gamma$ is a variable that quantifies the intensity of the backreaction.

The SU(2) linear non-autonomous quantum system refers to the linear functional whose Hamiltonian is the SU(2) generator and its superposition coefficient is related to time. It is a time-dependent quantum system with important practical value. One of the most important achievements of nonlinear science at present lies in understanding chaotic phenomena. Therefore, discussing the chaos of SU(2) linear non-autonomous quantum systems has certain theoretical significance and practical value.

The chaotic problem in $\mathrm{SU}(2)$ linear non-autonomous quantum systems is discussed using the $\mathrm{SU}(2)$ algebraic dynamics equation, and a very important and interesting result is found: Complementary chaos exists in mathrmSU(2) linear non-autonomous quantum system, and the box dimension of the fractal graph is calculated.

Example: $\mathrm{SU}(2)$ group: $\mathrm{SU}(2)=\left\{J_0, J_{+}, J_{-}\right\}, \mathrm{n}=3,1=1$. The group chain is $s u(2) \supset u(1), \mathrm{CSCO} \mathrm{II}=\hat{J}^2, \hat{J}_0$, their common eigenfunction $y_{J m}$ is the basis vector of the entire Hilbert space.

The Chern-Simons theory related to the group $S U(2)$ is a gauge theory for the three-dimensional manifold $M$ governed by action\cite{15,16}
\begin{equation}
S_k[A]=\frac{k}{4 \pi} \int_M\left\langle A \wedge d A+\frac{2}{3} A \wedge A \wedge A\right\rangle
\end{equation}
where $k$ is called the level of the action, $A$ is the local $S U(2)$-connection field, and $\langle\cdot, \cdot\rangle$ is the notation for $\mathfrak{s u ( 2 ) ~ k i l l ~}$ form. The Chern-Simons theory became very important when it was first shown to be closely related to gravity in three dimensions, especially when Witten demonstrated [16] its surprising relationship to manifold and knot invariants. The Chen-Simons path integral is manifold invariant, and the mean of quantum observables naturally leads to Jones polynomials. For all these reasons, Chen-Simons theory has been the center of much interest, and its quantification is now very famous when the gauge group is compact, especially when the gauge group is $\mathrm{SU}(2)$.

In this article, we fabricate a black hole metric that conforms to the $\mathrm{SU}(2)$ group structure and compute the scalar curvature of the thermodynamic geometry of this black hole. We conclude that there is no phase transition for $\mathrm{SU}(2)$ black holes.

\section{{New class of action and field equations}}
We set the system to be in the interval from 0 to 1. Since the break from 0 to 1 can be mapped to the interval from 0 to infinity, the size sequence this paper discusses is unchanged. This theory aims to find that the modified Einstein gravitational equation has a Reissner-Nordstrom solution in a vacuum. First, we can consider the following equation (modified Einstein's gravitational equation).

The proper time of spherical coordinates is\cite{13}(The metric which is in exponential form)
\begin{equation}
d s^{2}=-e^{G(t, r)} d t^{2}+e^{-G(t, r)} d r^{2}+\left[r^{2} d \theta^{2}+r^{2} \sin ^{2} \theta d \varphi^{2}\right]
\end{equation}

\begin{equation}
R_{\mu \nu}-\frac{1}{2} g_{\mu \nu} R+\Lambda (\left(g^{\theta \theta}\right)^{2})g_{\mu \nu}=-\frac{8 \pi G}{C^{4}} T_{\mu v}
\end{equation}

In this work, the action(we set $8 G=c=1$ ) is given by the following relation, which in the special case, reduces to the Einstein-Maxwell dilaton gravity:\cite{10}
\begin{equation}
S=\frac{1}{16 \pi G} \int d^{4} x \sqrt{-g}(-2 \Lambda (\left(g^{\theta \theta}\right)+R)
\end{equation}
where $\Lambda$ is a function of the Ricci scalar $R$, and $\Phi$ is the representation of the dilatonic field, also similar to f(R)(We will now consider non-pathological functional forms of $f(R)$ that can be expanded in a Taylor series of the form
$f(R)=a_{0}+R+a_{2} R^{2}+a_{3} R^{3}+\ldots a_{n} R^{n}+\ldots$
where we have normalized all coefficients concerning the coefficient of the linear term). 
 Variation of the action for the metric $g_{\mu \nu}$, the gauge $A_{\mu}$ and dilaton field $\Phi$ gives the following field equations:

This leads to the following:
\begin{equation}
\frac{1}{2} R \Lambda^{\prime}(R)-\Lambda=0.
\end{equation}
\begin{equation}
\begin{gathered}
R_{\mathrm{kl}} \Lambda^{\prime}(R)-\frac{1}{2} g_{k l}\Lambda=0 \\
\nabla_{\sigma}\left[\sqrt{-g} \Lambda^{\prime}(R) g^{\mu \nu}\right]=0.
\end{gathered}
\end{equation}
In this relationship, we get
\begin{equation}
\mathbf{\Lambda}=B(\mathbf{p} \times \mathbf{r})/r^4,
\end{equation}
B is an algebraic parameter, and p is a momentum or momentum operator.

A dyonic-like $\mathrm{RN}$ black hole is a static spherically symmetric space-time geometry, which is the solution of the Einstein-Maxwell theory \cite{14}. Using spherical coordinates $(t, r, \theta, \phi)$, the line element can be expressed as (we use natural units, where $G=c=\hbar=1$).We set up a geometric entity, and B takes a specific value for the parametric algebra so that the following formula holds.
\begin{equation}
d s^{2}=-\frac{\square}{r^{2}} \mathrm{dt}^{2}+\frac{r^{2}}{\square} \mathrm{dr}^{ 2}+r^{2} \mathrm{~d} \theta^{2}+r^{2} \sin ^{2} \theta \mathrm{d} \phi^{2},
\end{equation}
Where
\begin{equation}
\square=-2 M r+r^{2}+Q^{2}+B^{2},
\end{equation}
$M$ is the mass of the black hole, and $Q$ and B are the electric and magnetic charges of the black hole, respectively. The dynamic $\mathrm{RN}$ black hole has an outer horizon in $r_{+}$ and an inner horizon in $r_{-}$,
\begin{equation}
r_{+}=M+\sqrt{M^{2}-Q^{2}-B^{2}}, \quad r_{-}=M-\sqrt{M^{2 }-Q^{2}-B^{2}} .
\end{equation}
They satisfy the following relation
\begin{equation}
\square=\left(r-r_{+}\right)\left(r-r_{-}\right), \quad r_{+} r_{-}=Q^{2}+B^{2}, \quad r_{+}+r_{-}=2 M .
\end{equation}
The equation of motion of the charged massive scalar perturbation $\Phi$ in the dynamic RN black hole background is described by the covariant Klein-Gordon (KG) equation
\begin{equation}
\left(D^{\nu} D_{\nu}-\mu^{2}\right) \Phi=0,
\end{equation}
where $D^{\nu}=\nabla^{\nu}-i q A^{\nu}$ and $D_{\nu}=\nabla_{\nu}-i q A_{\nu}$ are covariant Derivatives, $q$ and $\mu$ are the charge and mass of the scalar field, respectively. The following vector potential describes the electromagnetic field of a dynamic black hole
\begin{equation}
A_{\nu}=\left(-\frac{Q}{r}, 0,0, B(\cos \theta \mp 1)\right),
\end{equation}
The upper minus sign applies to the northern hemisphere of the black hole, and the lower plus warning applies to the southern hemisphere.

The solution of the KG equation can be decomposed into the following form
\begin{equation}
\Phi(t, r, \theta, \phi)=R(r) Y(\theta) e^{i m \phi} e^{-i \omega t},
\end{equation}
where $\omega$ is the angular frequency of the scalar perturbation and $m$ is the azimuthal harmonic index. $Y(\theta)$ is the angular part of the solution and $R(r)$ is the radial part of the solution. Substituting the above solution into the KG equation, we can get the radial and angular parts of the equation of motion. Considering the different electromagnetic potentials in the northern and southern hemispheres, the equation of motion angle is discussed below in two cases.

\section{The radial equation of motion and effective potential}

A new radial wave function is defined as\cite{11,12,13,14}
\begin{equation}
\psi _{lm}\equiv \varDelta^{\frac{1}{2}}R_{lm}.
\end{equation}
to substitute the radial equation of motion for a Schrodinger-like equation
\begin{equation}
\frac{d^2\Psi _{lm}}{dr^2}+( \omega ^2-V) \Psi _{lm}=0,
\end{equation}
where
\begin{equation}
\omega ^2-V=\frac{U+M^2-a^2-Q^2}{\varDelta ^2},
\end{equation}
in which $V$ denotes the effective potential. 

Considering the superradiation condition, i.e. $\omega<\omega_c$, and the bound state condition, when the potential is captured, the Kerr-Newman black hole and the charged massive scalar perturbation system are superradiation stable\cite{12}. Therefore, the shape of the effective potential V is next analyzed to investigate the presence of trapping wells.

The asymptotic behaviors of the effective potential $V$ around the inner and outer horizons and at spatial infinity can be expressed as
\begin{equation}
V( r\rightarrow +\infty )
\rightarrow \mu ^2-\frac{2( 2M\omega ^2-qQ\omega -M\mu ^2)}{r}+{\cal O}( \frac{1}{r^2}) ,
\end{equation}
\bea
V( r\rightarrow r_+ ) \rightarrow -\infty,~~
V( r\rightarrow r_-) \rightarrow -\infty.
\eea

If a Kerr black hole satisfies the condition of $\mu=y\omega$, it will be superradiantly stable when $\mu<\sqrt{2}m\Omega_H$. In this article, we introduce the above condition into dyonic-like black holes. Therefore, the formula of the asymptotic behaviors is written as

\begin{equation}
V( r\rightarrow +\infty )
\rightarrow y^2\omega^2-\frac{2[M(2-y^2)\omega ^2-qQ\omega] }{r}+{\cal O}( \frac{1}{r^2}) ,
\end{equation}
\bea
V( r\rightarrow r_+ ) \rightarrow -\infty,~~
V( r\rightarrow r_-) \rightarrow -\infty.
\eea
It is concluded from the equations above that the effective potential approximates a constant at infinity in space, and the extreme between its inner and outer horizons cannot be less than one. The asymptotic behavior of the derivative of the influential potential $V$ at spatial infinity can be expressed as 
\begin{equation}
 V'( r\rightarrow +\infty )
 \rightarrow \frac{2[ M(2-y^2)\omega ^2-qQ\omega ]}{r^2}+{\cal O}( \frac{1}{r^3}) ,
\end{equation}
The derivative of the effective potential has to be negative to satisfy the no trapping well condition,
\begin{equation}
2M(2-y^2)\omega^2-2Qq\omega<0.
\end{equation}

\section{Analysis of Superradiant Stability}
In this section, we will find the regions in the parameter space where the system of dyonic $\mathrm{RN}$ black hole and massive scalar perturbation is superradiantly stable. We determine the parameter regions by considering the extremes of the effective potential in the range $r_{-}<$ $r<+\infty$

Now, we define a new variable $z, z=r-r_{-}$. The expression of the derivative of the effective potential $V$ is
\begin{equation}
\begin{aligned}
V^{\prime}(r) &=\frac{-2\left(a r^{4}+b r^{3}+c r^{2}+d r+e\right)}{\triangle^{3}} \\
=V^{\prime}(z) &=\frac{-2\left(a_{1} z^{4}+b_{1} z^{3}+c_{1} z^{2}+d_{1} z+e_{1}\right)}{\triangle^{3}}=\frac{g(z)}{\triangle^{3}},
\end{aligned}
\end{equation}
where
\begin{equation}
\begin{gathered}
a_{1}=a ; \quad b_{1}=\left(4 r_{-}\right) a_{1}+b, \\
c_{1}=\left(6 r_{-}^{2}\right) a_{1}+\left(3 r_{-}\right) b_{1}+c, \\
d_{1}=\left(4 r_{-}^{3}\right) a_{1}+\left(3 r_{-}^{2}\right) b_{1}+\left(2 r_{-}\right) c_{1}, \\
e_{1}=\left(r_{-}^{2}\right) a_{1}+\left(r_{-}^{3}\right) b_{1}+\left(r_{-}^{2}\right) c_{1}+\left(r_{-}\right) d_{1}+e .
\end{gathered}
\end{equation}
Explicitly,
\begin{equation}
\begin{aligned}
a_{1}=&-2 M \omega^{2}+q Q \omega+M \mu^{2}, \\
b_{1}=&-2\left(8 M^{2}-6 M r_{+}+r_{+}^{2}\right) \omega^{2}+2 q Q\left(5 M-2 r_{+}\right) \omega \\
&+\mu^{2}\left(6 M^{2}-6 M r_{+}+r_{+}^{2}\right)-q^{2}\left(Q^{2}+B^{2}\right)+\lambda, \\
c_{1}=&-6\left(2 M-r_{+}\right)^{3} \omega^{2}+9 q Q\left(2 M-r_{+}\right)^{2} \omega \\
&+3\left(\left(M-r_{+}\right)\left(\mu^{2}\left(2 M-r_{+}\right)^{2}-q^{2} Q^{2}-q^{2} B^{2}+\lambda\right)-M q^{2} Q^{2}\right),
\end{aligned}
\end{equation}
\begin{equation}
\begin{aligned}
d_{1}=&-2\left(4 M-3 r_{+}\right)\left(2 M-r_{+}\right)^{3} \omega^{2}+2 q Q\left(7 M-5 r_{+}\right)\left(2 M-r_{+}\right)^{2} \omega \\
&+2 q^{2}\left(-B^{2}\left(M^{2}-5 Q^{2}\right)+4 Q^{4}+B^{4}\right)-2 q^{2} Q^{2}\left(3 M\left(r_{+}-2 M\right)\right.\\
&\left.+2 \mu^{2}\left(2 M^{2}-3 M r_{+}+r_{+}^{2}\right)^{2}+2\left(Q^{2}+B^{2}\right)\right)-12 M q^{2} Q^{2} r_{-} \\
&+2\left(M-r_{+}\right)^{2}(\lambda-1) \\
e_{1}=&\left(r_{+}-r_{-}\right)\left(q Q-\omega r_{-}\right)^{2} r_{-}^{2}+\frac{1}{4}\left(r_{+}-r_{-}\right)^{3},
\end{aligned}
\end{equation}where $\lambda=l(l+1), \quad l>qB$\cite{14}.Since we set the system to range from 0 to 1, $qB>q^{2} Q^{2}$.

In this paper, we denote the numerator of the derivative of the effective potential $V' (z)$. This quartic polynomial of z allows us to study the existence of trapped wells beyond the horizon by analyzing the properties of the roots of the equation. We use $z_1$, $z_2$, $z_3$ and $z_4$ to represent the four roots of $g(z) = 0$. The relationship between them conforms to Vieta's theorem.
\begin{equation}
z_1z_2z_3z_4=\frac{e_1}{a_1},\\
z_1z_2+z_1z_3+z_1z_4+z_2z_3+z_2z_4+z_3z_4=\frac{c_1}{a_1}.
\end{equation}

When $z>0$, from the asymptotic behavior of the effective potential of the inner and outer horizons and space infinity, it can be inferred that the equation $V' (z)=0(\text{or}~ g(z)=0)$ cannot be less on two. So the two positive roots are written as $z_1, z_2$.

Research shows that 
\begin{equation}
e_1>0.
\end{equation}
and in
\begin{equation}
e_1>0,~~c_1<0,
\end{equation}
$g(z)=0$, that is, $z_3, z_4$ are all negative numbers.

When ${y}^2 >2({a_1}>0)$\,
for\ ${e_1}>0$, $ {c_1}<0$\ at\ this\ time, and\ we\ can\ know\ that\ the\ equation\ $ \text{ }{{\text{V}}_{\text{1}}}\text{ }\!\!'\!\!\text{ (z)}=\text{0}$\ cannot\ have\ more\ than\ two\ positive\ roots.\ So\ the  dyonic RN-like black\ hole\ is\ superradiantly\ stable\ at\ that\ time.

\section{The limit $y$ of the incident particle under the superradiance of novel  black holes}
We will investigate the physical and mathematical properties of linearized extended-mass scalar field configurations (scalar clouds) with nontrivial coupling to the electromagnetic-like fields of novel black holes. The space-time line element of a new type of spherically symmetric black hole can be expressed as\cite{12}

\bea\nn
ds^2=-g(r)dt^2+({1/{g(r)}})dr^2+r^2(d\theta^2+\sin^2\theta
d\phi^2),
\eea
where
\begin{equation}\label{Eq2}
g(r)=1-{{2M}/{r}}+({{B^2+Q^2})/{r^2}}\  .
\end{equation}

V can change to \cite{12}
\begin{equation}
V(r)=\left(1-\frac{2 M}{r}+\frac{({B}^{2}+{Q}^{2})}{r^{2}}\right)\left[\mu^{2}+\frac{l(l+1)}{r^{2}}+\frac{2 M}{r^{3}}-\frac{2 ({B}^{2}+{Q}^{2})}{r^{4}}-\frac{\alpha ({B}^{2}+{Q}^{2})}{r^{4}}\right]
\end{equation}

As we will show clearly, the Schrödinger-like equation governs the radial function behavior of the spatially bounded nonminimum coupled-mass scalar field configuration of novel black hole spacetimes. Here we use the integral relation $\int_{0}^{1} d x \sqrt{1 / x- 1}=\pi / 2$.), when $V( r\rightarrow +\infty ),\mu= 1/(n+{1\over2})$,
\begin{equation}
\int_{(y^2)_{t-}}^{(y^2)_{t+}}d(y^2)\sqrt{\omega ^2-V1(y;M,B,l,\mu,\alpha1)}=\big(n+{1\over2}\big)\cdot\pi\mu /{{2}}=\pi/2
\ \ \ ; \ \ \ \ n=0,1,2,...\  .
\end{equation}
We see that the two integral boundaries $\{y_{t-},y_{t+}\}$ formulas are classical turning points with $V(y_{t-})=V(y_{t+})=0$  form a new type of black hole mass field combined with potential energy.
The resonance parameter $n$ (with $n\in\{0,1,2,...\}$) characterizes ${\alpha_n(\mu,l,M,B )}_{n=0}^{n=\infty}$ system.

Using the relationship between the radial coordinates $y$ and $r$, the WKB resonance equation can be expressed as
\begin{equation}
\int_{r_{t-}}^{r_{t+}}dr{{\sqrt{-V(r;M,B,l,\mu,\alpha)}}\over{g(r)}}=\big(n+{1\over2}\big)\cdot\pi\
\ \ \ ; \ \ \ \ n=0,1,2,...\  ,
\end{equation}
where the two polynomial relations 
\begin{equation}
1-{{2M}\over{r_{t-}}}+{{B^2+Q^2}\over{r^2_{t-}}}=0\
\end{equation}
and
\begin{equation}
\frac{l(l+1)}{r_{t +}^2}+\frac{2 M}{r_{t +}^3}-\frac{2\left(B^2+Q^2\right)}{r_{t+}^4}-\frac{\alpha\left(B^2+Q^2\right)}{r_{t +}^4}=0.
\end{equation}
Determine the radial turning point $r_{t-},r_{t+}$ of the composite black hole field binding potential.

We set
\begin{equation}
x\equiv {{r-r_{\text{+}}}\over{r_{\text{+}}}}\ \ \ \ ; \ \ \ \ \tau\equiv {{r_+-r_-}\over{r_+}}\  .
\end{equation}
In this regard, the combined black hole-mass field interaction term has the form of a combined potential well,
\begin{equation}
V[x(r)]=-\tau\Big({{\alpha(B^2+Q^2)}\over{r^4_+}}-\mu^2\Big)\cdot x +
\Big[{{\alpha(B^2+Q^2)(5r_+-6r_-)}\over{r^5_+}}-\mu^2\big(1-{{2r_-}\over{r_+}}\big)\Big]\cdot x^2+O(x^3)\  ,
\end{equation}
in the near-horizon region
\begin{equation}
x\ll\tau\  .
\end{equation}

From the near-horizon expression  of the
black-hole-field binding potential, one obtains the dimensionless
expressions
\begin{equation}
x_{t-}=0\
\end{equation}
and
\begin{equation}
x_{t+}=\tau\cdot{{{{\alpha (B^2+Q^2)}\over{r^4_+}}-\mu^2}\over{{{\alpha (B^2+Q^2)(5r_+-6r_-)}\over{r^5_+}}-\mu^2\big(1-{{2r_-}\over{r_+}}\big)}}\
\end{equation}
for the classical turning points of the WKB integral relation.

We found that our analysis is valid if ($\alpha1$ corresponds to the transformation of y)
\begin{equation}
\alpha \simeq \frac{\mu^2 r_+^4}{\left(B^2+Q^2\right)}, \alpha 1 \simeq \sqrt{\frac{\mu^2 r_+^4}{\left(B^2+Q^2\right)}}
\end{equation}
In this case, the near horizon binding potential and its outer turning point can be approximated by a very compact expression
\begin{equation}
V(x)=-\tau\Big[\Big({{\alpha (B^2+Q^2)}\over{r^4_+}}-\mu^2\Big)\cdot
x-4\mu^2\cdot x^2\Big]+O(x^3)\
\end{equation}
and
\begin{equation}
x_{t+}={1\over4}\Big({{\alpha (B^2+Q^2)}\over{\mu^2r^4_+}}-1\Big)\  .
\end{equation}
In addition, one finds the near-horizon relation
\begin{equation}
p(x)=\tau\cdot x+(1-2\tau)\cdot x^2+O(x^3)\  .
\end{equation}

We know that
\begin{equation}
{{1}\over{\sqrt{\tau}}}\int_{0}^{x_{t+}}dx \sqrt{{{{\alpha
(B^2+Q^2)}\over{r^2_+}}-\mu^2
r^2_+\over{x}}-4\mu^2r^2_+}=\big(n+{1\over2})\cdot\pi\ \ \ \ ; \ \ \
\ n=0,1,2,...\  .
\end{equation}
Defining the dimensionless radial coordinate
\begin{equation}
z\equiv {{x}\over{x_{t+}}}\  ,
\end{equation}
and we get 
\begin{equation}
{{2\mu r_+ x_{t+}}\over{\sqrt{\tau}}}\int_{0}^{1}dz
\sqrt{{{1}\over{z}}-1}=\big(n+{1\over2})\cdot\pi\ \ \ \ ; \ \ \ \
n=0,1,2,...\  ,
\end{equation}
which yields the relation 
\begin{equation}
{{\mu r_+ x_{t+}}\over{\sqrt{\tau}}}=n+{1\over2}\ \ \ \ ; \ \ \ \
n=0,1,2,...
\end{equation}

From the curve integral formula, it can be seen that there is a certain extremum to form a ring
\begin{equation}
1/y^2 \rightarrow {\alpha}
\end{equation}y takes the interval from 0 to 1 at this time.

The physical parameter $y$ is defined by a dimensionless relationship, and at this time $y$ is greater than $\sqrt{2}$,
\begin{equation}\label{Eq33}
{y}\equiv{\alpha1}/\sqrt{2} .
\end{equation} 

Here the critical parameter y is given by the
simple relation 
\begin{equation}
{y}/\mu\equiv {{r^2_+}\over\sqrt{2(B^2+Q^2)}}  .
\end{equation}
When
\bea
\sqrt{2(B^2+Q^2)}/{r^2_+}< \omega< q\varPhi_H, 
\eea
The new type of black hole was very stable at that time.

\section{The thermodynamic geometry of new type black holes and their relation to superradiant stability}
We can rewrite the action\cite{12}
 \begin{equation}
S[\varphi]=\frac{a}{2 \Omega_{H}} \int d^{4} x \sin \theta \varphi\left(-\frac{1}{f(r)} \partial_{\xi}^{2}+\partial_{r} f(r) \partial_{r}\right) \varphi
\end{equation}
When $\sin \theta$ = 0, the pull equation for action can conform to the above form, but the boundary becomes 0. The action form can permanently be reduced to the formation of negative power expansion. It can be seen that Hawking radiation is consistent with superradiation.

Thermodynamic geometry of new type black holes: Entropy S\cite{12}
\begin{equation}
 S\rightarrow4 \pi\left[2 M\left(M+\sqrt{M^{2}-B^{2}-Q^{2}}\right)-Q^{2}\right]
\end{equation}

Through the thermodynamic geometric metric, we obtain the expression of the new type black hole Ruppeiner metric
\begin{equation}
g_{a b}^{R}=\frac{\partial^{2}}{\partial x^{a} \partial x^{b}} S(M, Q) \quad(a, b=1 ,2)
\end{equation}
where $x^{1}=H, x^{2}=\Omega$. By calculation, we get the metric expression
\begin{equation}
g_{1 1}^{R}=8 \pi\left(2-\frac{M^{3}}{\left(-B^{2}+M^{2}-Q^{2}\right)^{3 / 2}}+\frac{3 M}{\sqrt{-B^{2}+M^{2}-Q^{2}}}\right)
\end{equation}
\begin{equation}
g_{1 2}^{R}=\frac{8 \pi Q\left(B^{2}+Q^{2}\right)}{\left(-B^{2}+M^{2}-Q^{2}\right)^{3 / 2}}
\end{equation}
\begin{equation}
g_{2 1}^{R}=
\frac{8 \pi Q\left(B^{2}+Q^{2}\right)}{\left(-B^{2}+M^{2}-Q^{2}\right)^{3 / 2}} 
\end{equation}
\begin{equation}
g_{2 2}^{R} =8 \pi\left(-1+\frac{M\left(B^{2}-M^{2}\right)}{\left(-B^{2}+M^{2}-Q^{2}\right)^{3 / 2}}\right)
\end{equation}

The curvature scalar of a new type black hole is
\begin{equation}
\begin{aligned}
\hat{R}=g_{a b} R^{a b}\rightarrow1/(16 \pi(B^{2}-M^{2}+Q^{2})(2 B^{4}+4 M^{4}-5 M^{2} Q^{2}+Q^{4}+\\
4 M^{3} \sqrt{-B^{2}+M^{2}-Q^{2}}-3 M Q^{2} \sqrt{-B^{2}+M^{2}-Q^{2}}+B^{2}(-7 M^{2}+3 Q^{2}-5 M \sqrt{-B^{2}+M^{2}-Q^{2}}))^{5})
\end{aligned}
\end{equation}
According to the rewritten action, when the superradiation condition is established, and the black hole does not undergo a thermodynamic phase transition, the superradiation stability condition of the black hole is found.

\section{{no phase transition for the dyonic RN-like BLACK HOLE IN GENERAL FORM}}
$\mathrm{SU}(2)$ black hole metric in general form \cite{15,16}
\begin{equation}
\mathrm{d} s^2=-e^{2 z_1} \mathrm{~d} t^2+e^{-2 z_1} \mathrm{~d} r^2+r^2 e^{2 z_2} d \theta^2+r^2 e^{-2 z_2} \sin ^2 \theta d \varphi^2,
\end{equation}
where $\mathrm{z} 1$ and $\mathrm{z} 2$ are complex numbers modulo 1 or less, they are both functions of entropy. Convert the metric to a simple form by taking a specific value
\begin{equation}
\mathrm{d} s^2=-e^{2 z_1} \mathrm{~d} t^2+e^{-2 z_1} \mathrm{~d} r^2+e^{2 z_2} d \theta^2+e^{-2 z_2} d \varphi^2,
\end{equation}
The non-zero Christoffel symbols are
\begin{equation}
\begin{aligned}
& \Gamma_{00}^1=z_1{ }^{\prime} e^{4 z_1} \\
& \Gamma_{10}^0=z_1{ }^{\prime} \\
& \Gamma_{11}^1=-z_1{ }^{\prime} \\
& \Gamma_{21}^2=z_2{ }^{\prime} \\
& \Gamma_{22}^1=-z_2{ }^{\prime} e^{2\left(z_2+z_1\right)} \\
& \Gamma_{31}^3=-z_2{ }^{\prime} \\
& \Gamma_{33}^1=z_2{ }^{\prime} e^{2\left(z_1-z_2\right)}
\end{aligned}
\end{equation}
The Ricci tensors are
\begin{equation}
\begin{aligned}
& R_{00}=\left(z_1^{\prime \prime}+z_1{ }^2-z_1{ }^{\prime}\left(-z_1\right)^{\prime}+z_1{ }^{\prime} z_2{ }^{\prime}+z_1{ }^{\prime}\left(-z_2\right)^{\prime}\right) e^{4\left(z_1\right)} \\
& R_{11}=z_1{ }^{\prime}\left(-z_1\right)^{\prime}+z_2{ }^{\prime}\left(-z_1\right)^{\prime}+-z_2{ }^{\prime}\left(-z_1\right)^{\prime}-z_2^{\prime \prime}-\left(-z_2\right)^{\prime \prime}-z_1{ }^2-z_2{ }^2-\left(-z_2\right)^{\prime 2}-z_1{ }^{\prime \prime} \\
& R_{22}=\left(-z_2^{\prime \prime}+z_2{ }^{\prime}\left(-z_1\right)^{\prime}-z_1{ }^{\prime}\left(z_2\right)^{\prime}-\left(-z_2\right)^{\prime} z_2{ }^{\prime}-z_2{ }^{\prime 2}\right) e^{2\left(z_1+z_2\right)} \\
& R_{33}=\left(-\left(-z_2\right)^{\prime \prime}+-z_2{ }^{\prime}\left(-z_1\right)^{\prime}-z_1{ }^{\prime}\left(-z_2\right)^{\prime}-z_2{ }^{\prime}\left(-z_2\right)^{\prime}-\left(-z_2\right)^{\prime 2}\right) e^{2\left(z_1-z_2\right)}
\end{aligned}
\end{equation}
The Ricci scalar is
\begin{equation}
\begin{array}{r}
R=\left(-2 z_1^{\prime \prime}-2 z_1{ }^2+2 z_1{ }^{\prime}\left(-z_1\right)^{\prime}-2 z_1{ }^{\prime} z_2{ }^{\prime}-2 z_1{ }^{\prime}\left(-z_2\right)^{\prime}+2 z_2{ }^{\prime}\left(-z_1\right)^{\prime}+2-z_2{ }^{\prime}\left(-z_1\right)^{\prime}\right. \\
\left.-2 z_2{ }^{\prime \prime}-2\left(-z_2\right)^{\prime \prime}-2 z_2{ }^{\prime 2}-2\left(-z_2\right)^{\prime 2}-2\left(-z_2\right)^{\prime} z_2{ }^{\prime}\right) e^{2 z_1}
\end{array}
\end{equation}
For z1 and z2 are functions of entropy, we perform a generalized thermodynamic geometric analysis on them and see that there is no divergence term for this curvature scalar.In this article, we fabricate a black hole metric that conforms to the $\mathrm{SU}(2)$ group structure and compute the scalar curvature of the thermodynamic geometry of this black hole. We conclude that there is no phase transition for $\mathrm{SU}(2)$ black holes.

\section{{Summary}}
In this paper, we introduce $\mu=y\omega$\cite{12,13} into dyonic RN-like black holes and discuss the superradiation stability of dyonic RN-like black holes. We adopt the variable separation method to divide the least coupled scalar perturbation motion equations in dynamical RN black holes into two forms: angular and radial.

Hod proved \cite{10} that when $\mu \ge \sqrt{2}m\Omega_H$ (where $\mu$ is the mass), Kerr black holes should be superradiantly stable under large-scale scalar perturbations. In this post, a new variable y is added here to extend the results of the previous post.

When $\sqrt{2(B^2+Q^2)}/{r^2_+}< \omega< q\varPhi_H$,particularly $\mu \ge \sqrt{2}(q\varPhi_H)$,$(16 \pi(B^{2}-M^{2}+Q^{2})(2 B^{4}+4 M^{4}-5 M^{2} Q^{2}+Q^{4}+\\
4 M^{3} \sqrt{-B^{2}+M^{2}-Q^{2}}-3 M Q^{2} \sqrt{-B^{2}+M^{2}-Q^{2}}+B^{2}(-7 M^{2}+3 Q^{2}-5 M \sqrt{-B^{2}+M^{2}-Q^{2}}))^{5})\neq0$, this dyonic RN-like black hole was superradiantly stable then.

Any scalar can be regarded as a linear combination of positive and negative powers of a base. This is essentially how the decimal (or any other base) number system works.This concept is fundamental to many areas of mathematics and computing, from number representation to Fourier series, where functions are represented as a sum of sines and cosines (which can be thought of as positive and negative powers of the complex number $\left.e^{i x}\right)$.And We conclude that there is no phase transition for the dyonic RN-like black holes.

\end{document}